\documentclass[aps,prl,numerical,linenumbers,superscriptaddress,showpacs,floatfix,reprint]{revtex4-1}
\usepackage[dvipdfm]{graphicx}
\usepackage{amsmath,amssymb,amsfonts}
\usepackage{color}

\newcommand{\be}{\begin{eqnarray}}
\newcommand{\ee}{\end{eqnarray}}

\def\v2{\mbox{$v_2$}}

\newcommand{\mean}[1]{\left\langle #1 \right\rangle}

\begin{document}
\title{Scaling patterns for azimuthal anisotropy in Pb+Pb collisions \\at $\sqrt{s_{NN}}=2.76$~TeV:
Further constraints on transport coefficients}
%
\author{ Roy~A.~Lacey} 
\email[E-mail: ]{Roy.Lacey@Stonybrook.edu}
\author{ N.~N.~Ajitanand} 
\author{ J.~M.~Alexander}
\author{ J.~Jia}
\author{A.~Taranenko}
%
\affiliation{Department of Chemistry, 
Stony Brook University, \\
Stony Brook, NY, 11794-3400, USA}


\date{\today}
\begin{abstract}

	Azimuthal anisotropy measurements for charged hadrons, characterized by the second 
order Fourier coefficient $v_2$, are used to investigate the path length ($L$) and transverse 
momentum ($p_T$) dependent jet quenching patterns of the QCD medium produced in Pb+Pb collisions 
at $\sqrt{s_{NN}}=2.76$\,TeV. $v_2$ shows a linear decrease as $1/\sqrt{p_T}$ and a linear 
increase with the medium path length difference ($\Delta L$) in- and out of the $\Psi_2$ event plane.
These patterns compliment a prior observation of the scaling of jet quenching ($R_{\rm AA}$) measurements. 
Together, they suggest that radiative parton energy loss is a dominant mechanism for 
jet suppression, and $v_2$ stems from the difference in the parton propagation length $\Delta L$.
An estimate of the transport coefficient $\hat{q}$, gives a value comparable to  
that obtained in a prior study of the scaling properties of $R_{\rm AA}$. 
These results suggest that high-$p_T$ azimuthal anisotropy measurements provide 
strong constraints for delineating the mechanism(s) for parton energy loss, 
as well as for reliable extraction of $\hat{q}$.

\end{abstract}
\maketitle


Ultrarelativistic heavy ion collisions can produce a high energy-density 
plasma of quarks and gluons (QGP) \cite{Shuryak:1978ij}. 
Full characterization of the transport properties of this plasma,
is a central objective of current research at both the Relativistic Heavy Ion Collider (RHIC)
and the Large Hadron Collider (LHC). 
A key ingredient for such a characterization is a full understanding of the 
mechanism by which hard scattered partons interact and lose energy in the QGP, prior 
to their fragmentation into topologically aligned high-$p_T$ 
hadrons or jets \cite{Gyulassy:1993hr}. 
This energy loss manifests as a suppression of hadron yields \cite{Adcox:2001jp} 
-- termed ``jet quenching'' -- which depends on the momenta of the partons and 
the path length for their propagation through 
the QGP \cite{Gyulassy:1993hr,Dokshitzer:2001zm,Lacey:2009ps,Lacey:2012bg}. 

Such a suppression is routinely quantified with the measured hadron yields in 
A+A and p+p collisions, via the nuclear modification 
factor ($R_{\rm AA}$) \cite{Adcox:2001jp,Adler:2002xw};
\begin{equation}
   R_{\rm AA}(p_T) = \frac{1/{N_{\rm evt}} dN/dydp_{\rm T}}{\mean{T_{\rm AA}} d\sigma_{pp}/dydp_{\rm T}}, 
\label{Eq.1}
\end{equation}
where $\sigma_{pp}$ is the particle production cross section in p+p collisions 
and $\mean{T_{\rm AA}}$ is the nuclear thickness function
averaged over the impact parameter ($\mathbf{b}$) range associated with a given 
centrality selection
\begin{equation}
\langle T_{AA}\rangle\equiv
\frac {\int T_{AA}(\mathbf{b})\, d\mathbf{b} }{\int (1- e^{-\sigma_{pp}^{inel}\, T_{AA}(\mathbf{b})})\, d\mathbf{b}}.
\label{Eq.2}
\end{equation}
The average number of nucleon-nucleon collisions, 
$\langle N_{coll}\rangle=\sigma_{pp}^{inel} \langle T_{AA}\rangle$,
is usually obtained via a Monte-Carlo Glauber-based 
model calculation \cite{Miller:2007ri,Alver:2006wh}.
Detailed measurements of the centrality and $p_T$ dependence of $R_{\rm AA}$ are 
key to ongoing efforts to delineate the transport properties of 
the QGP produced in heavy ion collisions at both the LHC and RHIC \cite{Qin:2007rn,Bass:2008rv,Lacey:2009ps,
Sharma:2009hn,CasalderreySolana:2010eh,Qin:2010mn,Renk:2011gj,Chen:2011vt,
Majumder:2011uk,Zakharov:2011ws,Betz:2012qq,Lacey:2012bg}. 

Differential measurements of the azimuthal anisotropy of high-$p_T$ hadrons also provide 
an indispensable probe to study jet quenching. Here, the operational 
ansatz is that the partons which traverse the QGP medium in the direction parallel (perpendicular) 
to the event plane result in less (more) suppression due to the shorter (longer) parton propagation 
lengths \cite{Gyulassy:2000gk,Wang:2000fq,Liao:2008dk,Jia:2011pi}. 
The resulting anisotropy can be characterized via Fourier decomposition of the measured azimuthal 
distribution;
\begin{equation}
\frac{dN}{d\phi} \propto \left(
1 + \sum_{n=1} 2 \, v_n \, \cos(n[\phi-\Psi_n]) \right),  
\label{Eq.3}
\end{equation}
where $v_n = \mean{ \cos(n[\phi - \Psi_{n}])}, n=1,2,3,...$
and the $\Psi_n$ are the generalized participant event
planes at all orders for each event.
Characterization can also be made via the pair-wise distribution 
in the azimuthal angle difference ($\Delta\phi =\phi_1 - \phi_2$) between 
particles \cite{Lacey:2001va,Adcox:2002ms,Aad:2012bu};
\begin{equation}
\frac{dN^{\text{pairs}}}{d\Delta\phi} \propto \left( 1 + \sum\limits_{n = 1} 
2v_n^a(p_T^a)v_n^b(p_T^b)\cos(n\Delta\phi) \right).
\label{Eq.4}
\end{equation}
%

%
\begin{figure*}[t]
\includegraphics[width=1.0\linewidth]{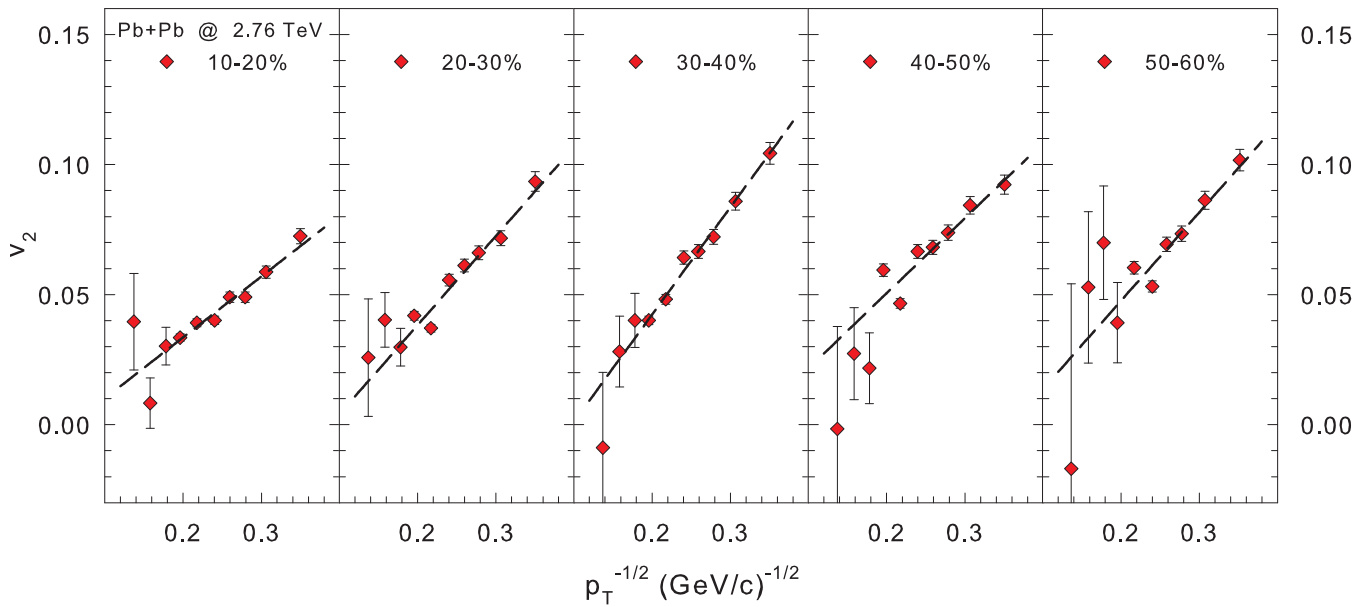}
\caption{ (Color online) $v_2(p_T)$ vs. $1/\sqrt{p_T}$ for several centrality 
selections as indicated. Error bars are statistical only. 
The data are taken from Refs. \cite{CMS_Prelim_v2,ATLAS:2011ah}. 
The dashed curve in each panel is a linear fit to the data.
}
\label{Fig1}
\end{figure*}

	In earlier work \cite{Lacey:2009ps,Lacey:2012bg}, we have investigated the 
$p_T$ and path length ($L$) dependence of jet quenching via the scaling properties 
of $R_{\rm AA}(p_T, L)$. The observed scaling indicated a decrease of $R_{\rm AA}(p_T, L)$ with $L$ 
and an increase of $R_{\rm AA}(p_T, L)$ with $1/\sqrt{p_T}$. These trends were shown to 
be compatible with an energy loss mechanism dominated by medium induced gluon 
radiation \cite{Dokshitzer:2001zm};
\begin{eqnarray}
R_{\rm AA}(p_T,L) \simeq \exp \left[- {2 \alpha_s C_F \over \sqrt{\pi}}\ 
L\,\sqrt{\hat{q}\frac{{\cal{L}}}{p_T}}\, \right] \nonumber \\
{\cal{L}} \equiv \frac{d}{d\ln p_T} 
\ln \left[ {d \sigma_{pp} \over d p_{T}^2}( p_{T})\right], 
\label{eq:DK1}	
\end{eqnarray}
where $\alpha_s$ is the strong coupling constant, $C_F$ is the color factor 
and $\hat{q}$ is the transport coefficient which characterizes the squared average 
transverse momentum exchange [per unit path length] between the medium 
and the parton. An estimate of $\hat{q}$ was also extracted from the scaling curves.
If jet quenching serves as a driver for azimuthal anisotropy, scaling as a function
of $p_T$ and $\Delta L = L_y - L_x$ (the difference between the out-of-plane ($y$) and in-plane ($x$) 
path lengths) might be expected. Thus, it is important to investigate whether 
azimuthal anisotropy measurements show complimentary scaling patterns which reflect 
the underlying energy loss mechanism suggested by the $R_{\rm AA}$ measurements.

	In this work, we use high-$p_T$ $v_2$ data to search for these scaling patterns,  
with an eye toward an independent estimate of $\hat{q}$. The observation of such 
patterns could also serve as a confirmation that at high-$p_T$, $R_{\rm AA}$ and $v_2$ 
stem from the same energy loss mechanism, and this mechanism is dominated 
by medium induced gluon radiation. 

	The high-$p_T$ ($p_T \agt 8$ GeV/$c$, $1 < \left|\eta \right| < 2 $) measurements employed in our search were  
recently reported for charged hadrons by the CMS and ATLAS collaborations \cite{CMS_Prelim_v2,ATLAS:2011ah}. 
These data indicate an increase of $v_2(p_T)$ from central to mid-central collisions, as might be 
expected from an increase in $\Delta L$ as collisions become more peripheral. They also indicate 
a characteristic decrease of $v_2$ with $p_T$, suggestive of a $1/\sqrt{p_T}$ dependence.
These key features are important to the scaling search discussed below.

To facilitate comparisons to our earlier scaling analysis of $R_{\rm AA}$ data,  
we use the transverse size of the system $\bar{R}$ as an estimate for the path 
length $L$, as was done in our earlier analyses \cite{Lacey:2009ps,Lacey:2012bg}. 
A Monte-Carlo Glauber-based model calculation \cite{Miller:2007ri,Alver:2006wh} 
was used to evaluate the values for $\bar{R}$ and the eccentricity $\varepsilon$ in Pb+Pb 
collisions as follows.
For each centrality selection, the number of participant nucleons 
$N_{\rm part}$, was first estimated. Subsequently, $\bar{R}$ and the eccentricity $\varepsilon$, were determined 
from the distribution of these nucleons in the transverse ($x,y$) plane as:
%
$
{1}/{\bar{R}}~=~\sqrt{\left(\frac{1}{\sigma_x^2}+\frac{1}{\sigma_y^2}\right)},
$
%
and $\varepsilon = \frac{\sigma_y^2 - \sigma_x^2}{\sigma_y^2 + \sigma_x^2}$,
where $\sigma_x$ and $\sigma_y$ are the respective root-mean-square widths of
the density distributions. We use the estimate 
$\Delta L \equiv L_y - L_x = \varepsilon (L_x+L_y) \sim \varepsilon \bar{R}$ for the path 
length difference. Note that $L_y = (\sigma_x \sqrt{1 + \varepsilon} )/(1 - \varepsilon )$
and $L_x = (\sigma_x \sqrt{1 + \varepsilon} )/(1 + \varepsilon )$.
For these calculations, the initial entropy profile in the transverse
plane was assumed to be proportional to a linear combination
of the number density of participants and binary collisions \cite{Hirano:2009ah,Lacey:2009xx}.
The latter assures that the entropy density weighting used, is constrained by the Pb+Pb hadron 
multiplicity measurements \cite{Chatrchyan:2011pb}. 
Averaging for each centrality, was performed over the configurations generated in the 
simulated collisions.
\begin{figure}[h]
\includegraphics[width=1.0\linewidth]{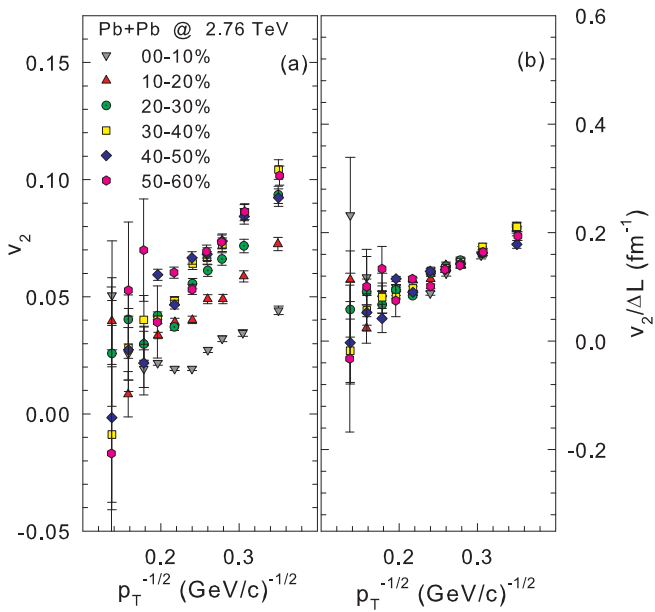}
\caption{ (Color online) (a) $v_2(p_T)$ vs. $1/\sqrt{p_T}$ for the centrality 
selections indicated. (b) $v_2(p_T)/\Delta L$ vs. $1/\sqrt{p_T}$ for the same centrality 
selections. Error bars are statistical only.
} 
\label{Fig2}
\end{figure}
\begin{figure*}
\includegraphics[width=0.95\linewidth]{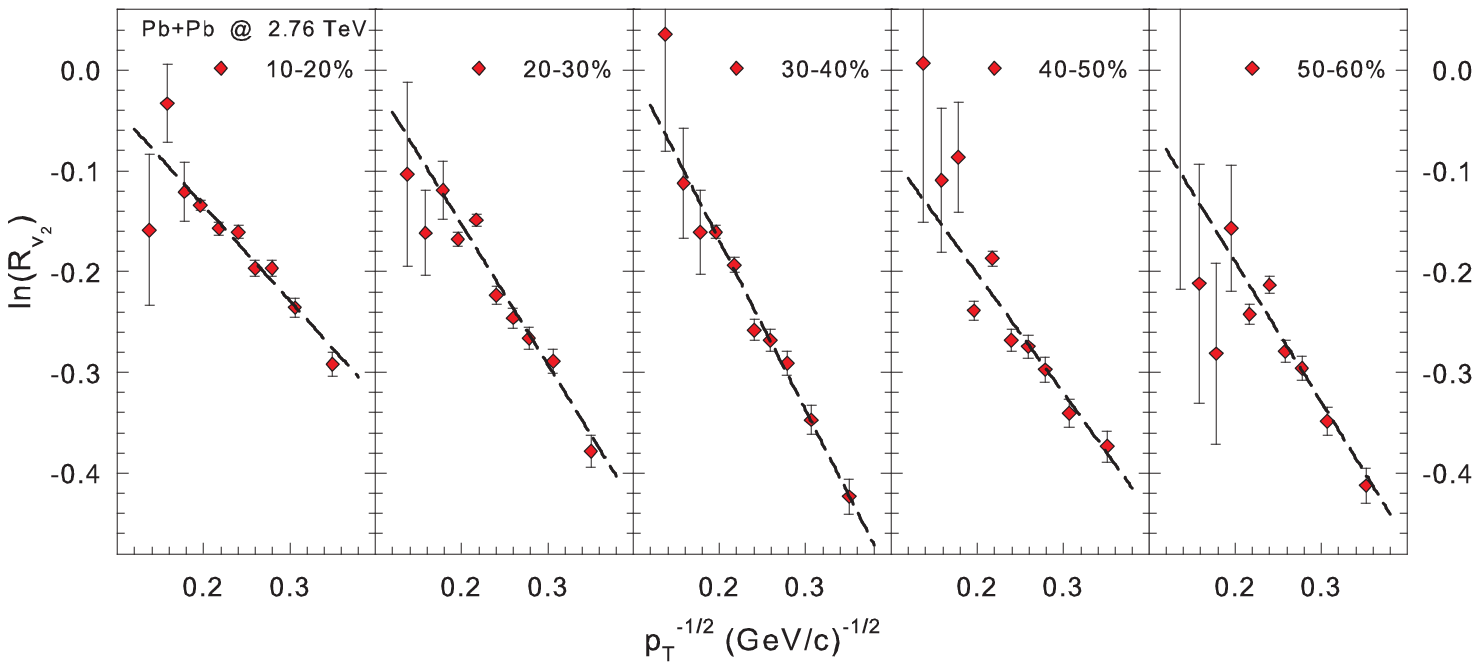}
\caption{(Color online) $\ln\left[R_{v_2}(p_T,\Delta L)\right]$ vs. $1/\sqrt{p_T}$  
for several centrality selections as indicated. Error bars are statistical only. 
The dashed curve in each panel shows a linear fit to the data (see text).
} 
\label{Fig3}
\end{figure*}

	Figure \ref{Fig1} shows the plots of $v_2(p_T)$ vs. $1/\sqrt{p_T}$ for several centrality 
selections as indicated. The dashed curves which represent a linear fit, indicates that within errors,
$v_2(p_T)$ decreases as $1/\sqrt{p_T}$. This trend is opposite to the trend for $R_{\rm AA}(p_T)$, and 
is to be expected if the anisotropy characterized by $v_2(p_T)$ stems 
from jet quenching (cf. Eq.~\ref{eq:DK1}), {\em i.e.} an increase in $R_{\rm AA}(p_T)$ results in 
a corresponding decrease in $v_2(p_T)$.	

	The combined effects of $1/\sqrt{p_T}$ and $\Delta L$ scaling are 
demonstrated in Fig.~\ref{Fig2}. The left panel (a) of the figure shows the same 
linear dependence on $1/\sqrt{p_T}$ evidenced in Fig.~\ref{Fig1}, but with a different 
magnitude for each of the centrality selections indicated. The right panel (b) shows that,  
when the same data [shown in (a)] is scaled by $\Delta L$, a single curve is obtained.
The implied linear increase of $v_2(p_T)$ with $\Delta L$ is complimentary to the previously 
observed $L$ dependence of jet quenching  \cite{Lacey:2009ps,Lacey:2012bg}. That is, 
an increase in the effective path length $L$ ($\Delta L$) leads to more quenching (anisotropy).  
An extrapolation to higher values of $p_T$, of a linear fit to the data in Fig.~\ref{Fig2}(b), 
suggests that the anisotropy associated with jet quenching is negligible 
($v_2 \sim 0$) for $p_T \agt 100$~GeV. This is probably due to the relatively small 
magnitudes of $\Delta L$. Note that for a fixed value of $p_T$, $\ln(R_{\rm AA}(p_T,L))$
shows a linear dependence on $L$ \cite{Lacey:2009ps,Lacey:2012bg}.

	For a given centrality, the azimuthal angle dependence of jet 
quenching [relative to the $\Psi_2$ event plane] $R_{\rm AA}(\Delta\phi,p_T)$, 
is related to $v_2(p_T)$. This stems from the fact that the number of particles emitted relative 
to $\Psi_2$, $N(\Delta\phi,p_T) \propto [1+2v_2(p_T)\cos(2\Delta\phi)]$. 
The anisotropy factor 
\begin{equation}
 R_{v_2}(p_T,\Delta L) = \frac{R_{\rm AA}(90^o,p_T)}{R_{\rm AA}(0^0,p_T)} = \frac{1-2v_2(p_T)}{1+2v_2(p_T)}, 
\label{Eq.5}
\end{equation}
({\em i.e} the ratio of the out-of-plane yield ($\Delta\phi=90^o$) to in-plane  yield ($\Delta\phi=0^o$)), 
quantifies the magnitude of the quenching for path length difference $\Delta L$.
Therefore, the values of $R_{v_2}(p_T,\Delta L)$ can be used in concert with 
Eq.~\ref{eq:DK1} to extract an estimate of $\hat{q}$. 
 
	To facilitate this estimate we first plot $\ln(R_{v_2}(p_T))$ vs. $1/\sqrt{p_T}$ for 
each centrality selection, as shown in Fig.~\ref{Fig3}. The dashed curve in each panel 
of the figure, represents a linear fit to the data; they show the expected linear 
dependence on $1/\sqrt{p_T}$ predicted by Eq.~\ref{eq:DK1}. 
The slopes ${S_{p_T}}$ of these curves encode the magnitude of both $\alpha_s$ and $\hat{q}$.

\begin{figure}[t]
\includegraphics[width=0.85\linewidth]{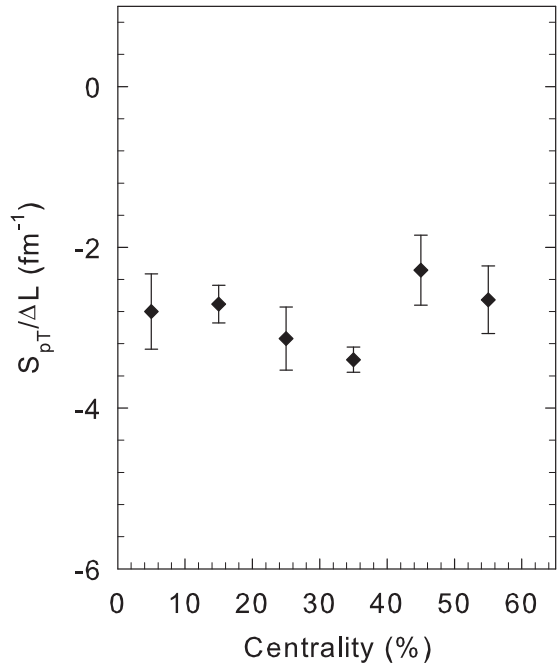}
\caption{(Color online) Centrality dependence of $S_{p_T}/\Delta L$, see text. 
The slopes $S_{p_T}$ are obtained from the the linear fits shown in Fig. \ref{Fig3}.
} 
\label{Fig4}
\end{figure}

	For a given medium (fixed $\left<\hat{q}\right>$) Eq.~\ref{eq:DK1} suggests that the ratio 
${S_{p_T}}/{\Delta L}$ should be independent of the collision centrality.  
This independence is indicated by the relatively flat centrality dependence for ${S_{p_T}}/{\Delta L}$, 
shown in Fig. \ref{Fig4}. This observation serves as a further 
validation of Eq.~\ref{eq:DK1}, so we use the average value 
of these ratios $\sim 3.0 \pm 0.3$ GeV$^{1/2}$/fm in concert with Eq.~\ref{eq:DK1}, to obtain the  
estimate $\hat{q}_{\rm LHC} \approx 0.47 \pm 0.09$~GeV$^2$/fm with values of 
$\alpha_s = 0.3$ \cite{Bass:2008rv}, $C_F = 9/4$ \cite{cfactor,Dokshitzer:2001zm} 
and ${\cal{L}} = n = 6.7$ \cite{ppPower}. This estimate of $\hat{q}_{\rm LHC}$, which can 
be interpreted as a space-time average, 
is similar to our earlier estimate $\hat{q}_{\rm LHC} \approx 0.56 \pm 0.05$~GeV$^2$/fm
from scaled $R_{\rm AA}$ data, evaluated with the same values for $C_F$ and $\alpha_s$ \cite{Lacey:2012bg}. 
We conclude that radiative parton energy loss drives jet suppression and  
a collateral azimuthal anisotropy ($v_2$) develops, due to the difference in the 
in-medium parton propagation length ($\Delta L$) in- and out of the $\Psi_2$ event plane.

	In summary, we have performed scaling tests on the $v_2$ values obtained 
from azimuthal anisotropy measurements of high-$p_T$ charged hadrons 
in Pb+Pb collisions at $\sqrt{s_{NN}} = 2.76$ TeV. 
$v_2$ shows a linear decrease as $1/\sqrt{p_T}$ and a linear increase with the 
medium path length difference ($\Delta L$) in the directions parallel and perpendicular 
to the $\Psi_2$ event plane.
These patterns, which are similar to the scaling patterns for jet quenching  
measurements ($\ln(R_{\rm AA})$), confirm the $1/\sqrt{p_T}$ dependence, as well as 
the linear dependence on path length predicted by Dokshitzer and Kharzeev for 
jet suppression dominated by the mechanism of medium-induced 
gluon radiation in a hot and dense QGP. These observations also suggest that, 
at high-$p_T$, $v_2$ stems from jet quenching, and is a direct consequence of 
the difference in the parton propagation length $\Delta L$.
A simple estimate of the transport coefficient $\hat{q}$ from the scaled $v_2$ data, 
gives a value which is similar to the value obtained in a prior study of the 
scaling properties of $R_{\rm AA}$ \cite{Lacey:2012bg}. 
These results confirm that high-$p_T$ azimuthal anisotropy measurements, provide
strong additional constraints for delineating the mechanism(s) for parton energy loss, 
as well as for reliable extraction of $\hat{q}$.

\section*{Acknowledgments}
This work was supported by the US DOE under contract DE-FG02-87ER40331.A008.
 

\bibliography{LHC_highpT-v2_Refs}

\begin{thebibliography}{34}
\bibitem [{cfa()}]{cfactor}%
  \BibitemOpen
  \href@noop {} {}\bibinfo {note} {Here, it is assumed that charged hadrons are
  dominantly produced via the fragmentation of gluons.}\BibitemShut {Stop}%
\end{thebibliography}
\end{document}